\documentclass[fleqn]{2017SCGE}
\usepackage{color}
\begin{document}

\ensubject{subject}

%%%%%%%%%%%%%%%%%%%%%%%%%%%%%%%%%%%%%%%%%%%%%%%%%%%%%%%
%%% Authors do not modify the information below
%%% ????????????????
%%% ??????????, ????????????{}, ???????????????????
%Letter to the Editor??Article%??????
\ArticleType{Article}%??Article
\SpecialTopic{SPECIAL TOPIC: }%???????
\Year{2017}
\Month{January}
\Vol{60}
\No{1}
\DOI{10.1007/s11432-016-0037-0}
\ArtNo{000000}
\ReceiveDate{January 11, 2016}
\AcceptDate{April 6, 2016}
%\OnlineDate{January 1, 2016}
%%%%%%%%%%%%%%%%%%%%%%%%%%%%%%%%%%%%%%%%%%%%%%%%%%%%%%%

%%% title: ????
%%%   \title{title}{title for citation}
\title{Multi-objective optimization in quantum parameter estimation}{Multi-objective optimization in quantum parameter estimation}

%%% Corresponding author: ???????
%%%   \author[number]{Full name}{{email@xxx.com}}
%%% General author: ???????
%%%   \author[number]{Full name}{}
\author[]{Beili Gong}{}%
\author[]{Wei Cui}{{aucuiwei@scut.edu.cn}}
%\author[1]{Zhongxing ZHANG}{}%\protect\\
%\author[1]{Liyuan LIU}{}%
%\author[1]{Nanjian WU}{}

%%% Author information for page head. ?¨¹?§Ö????????
%%% ??????????????, ??????????author???
\AuthorMark{Gong B}%\authorcr????????

%%% Authors for citation. ????????§Ö????????
%%% ??????????????, ??????????author???
\AuthorCitation{Gong B, Cui W}

%%% Address. ???
%%%   \address[number]{Address, City {\rm Postcode}, Country}
\address[]{School of Automation Science and Engineering, South China University of Technology, Guangzhou {\rm 510641}, China}
%\address[2]{For example: Institute of Mechanics, Chinese Academy of Sciences, Beijing 100190, China}

%\contributions{}%????????

%%% Abstract. ??
\abstract{We investigate quantum parameter estimation based on linear and Kerr-type nonlinear controls in an open quantum system, and consider the dissipation rate as an unknown parameter. We show that while the precision of parameter estimation is improved, it usually introduces a significant deformation to the system state. Moreover, we propose a multi-objective model to optimize the two conflicting objectives: (1) maximizing the Fisher information, improving the parameter estimation precision, and (2) minimizing the deformation of the system state, which maintains its fidelity. Finally, simulations of a simplified $\varepsilon$-constrained model demonstrate the feasibility of the Hamiltonian control in improving the precision of the quantum parameter estimation.}%

%%% Keywords. ?????


\maketitle

%\tableofcontents%?????

%%%%%%%%%%%%%%%%%%%%%%%%%%%%%%%%%%%%%%%%%%%%%%%%%%%%%%%
%%% The main text. ???????
%???????????????????\cref{fig1}
%\twocolumn\onecolumn
%%%%%%%%%%%%%%%%%%%%%%%%%%%%%%%%%%%%%%%%%%%%%%%%%%%%%%%
\begin{multicols}{2}
\section{Introduction}
Quantum information processing and quantum control often require accurate information of the parameters of the Hamiltonian system, the surrounding environment, the coupling and measurement strengths, and so on. However, due to the inevitable randomness of the quantum measurement, and to the fact that several quantities of interest cannot even be associated to proper quantum observables, quantum parameter estimation  \cite{Helstrom:1976,Wiseman:2010,Holevo:2011,Wangh:2016} has become a fundamental problem in quantum science and technology. Strictly speaking, this problem arises most often in gravitational-wave experiments \cite{Cui:2017,Li:2018,Hu:2017}, and the maximum sensitivity for the conventional continuous monitoring of the position of the probe mass is given by the standard quantum limit (SQL) for the sensitivity of the mass to the classical force \cite{Braginsky:1992}. However, Caves \cite{Caves:1981} showed that with the help of the squeezed state technique, quantum mechanical systems can achieve greater sensitivity over the SQL. Since then, scientists and engineers have developed various quantum technologies \cite{Giovannetti:2004,Giovannetti:2006,Zwierz:2010,Pesse:2007,Lovett:2013,Jin:2017,Cheng:2014,Zhang:2015} to improve the accuracy of a wide variety of quantum measurements. Theoretically the ultimate precision limit is the Heisenberg limit (HL), which relies on the unitarity of the time evolution.
When a quantum state is used as a probe and an optimization procedure is involved, a quantum version of the Cram\'{e}r$--$Rao inequality \cite{Braunstein:1994, Paris:2009,Jacobs:2014} can be established. In general, the Cram\'{e}r$--$Rao bound can be applied to any parameter estimation problem, with the quantum Fisher information (QFI) \cite{Lu:2010,Wang:2016,Yuan:2015} as the upper bound on the precision of the parameter estimation.

Contrary to closed quantum systems, quantum probes would inevitably interact with the surrounding environment. For these realistic conditions, the dissipative Cram\'{e}r$--$Rao bound has been derived, and the estimation accuracy remarkably depends on the underlying dynamical map with a semigroup property \cite{Alipour:2014,Fei:2016,Ma:2016}. Due to the profound nature of the theory of open quantum systems \cite{Breuer:2002}, improving a quantum estimation problem in the presence of noise attracts fundamental interest \cite{Escher:2011,Tsang:2013,Nair:2016,Molmer:2013,Molmer:2016}. In particular, much attention has been devoted to the estimation of the noisy frequency and loss parameters \cite{Smirne:2016,Spedalieri:2016,Rossi:2016}.
For example, in a recent work \cite{Rossi:2016}, it is shown that the estimation  accuracy of the rate of loss can be improved by a Kerr-type nonlinear Hamiltonian.
However, due to the dissipative evolution of the quantum system, the coherence of the initial state can be quickly damaged and the Gaussian input becomes a set of non-Gaussian states. Thus, the global optimization of these two conflicting objectives (1. maximizing the Fisher information, which improves the parameter estimation precision; 2. minimizing the deformation on the system state that maintains fidelity) is of great importance in the practical application of the quantum parameter estimation method.

In our work, we study the quantum parameter estimation based on the linear and Kerr-type nonlinear controls in the Hamiltonian of the system. In particular, the estimation of the dissipation rate of a quantum master equation has been considered. Using a pure state approximation, we obtain the QFI in analytical form. We verify the validity of this approximation by comparing the approximate QFI with the exact one with various controls.  We show that while we improve the precision of parameter estimation, this usually induces significant deformation on the system state. Moreover, we propose a multi-objective model \cite{Ehrgott:2006,Miettinen:1999} that can optimize the two conflicting objectives (QFI and fidelity) simultaneously.  Finally, simulations of a simplified $\varepsilon$-constrained model demonstrate the feasibility of the Hamiltonian control in the quantum parameter estimation.

The paper is organized as follows: in {\color{red}Sec.~\ref{sec:The_model}}, based on a simple quantum master equation, a detailed analysis of its solutions by pure state approximation is given. In {\color{red}Sec.~\ref{sec:Quantum_parameter_estimation}}, we calculate the quantum Fisher information with various control Hamiltonians.
 In {\color{red}Sec.~\ref{sec:Multi-objective_optimization}}, we study the trade-off between the QFI and the quantum fidelity by the multi-objective optimization theory.  We summarize our conclusions in {\color{red}Sec.~\ref{sec:Conclusion}}.

\section{The model}\label{sec:The_model}
In reality, quantum systems are inevitably interact with the environment, and the evolution of the system is described by a master equation \cite{Breuer:2002}. The simplest master equation describing amplitude dissipation is
\begin{equation}\label{Equ:Lindblad_Master_Equ}
\frac{{d\rho }}{{dt}} = \gamma \left( {a\rho {a^\dag } - \frac{1}{2}{a^\dag }a\rho  - \frac{1}{2}\rho {a^\dag }a} \right),
\end{equation}
where $\gamma $ is the dissipation rate of a bosonic channel, and $a$ is the annihilation operator. We consider $\gamma$ as an unknown parameter, which required to be estimated. The precision of the parameter estimation is always given by the quantum Cram\'{e}r$--$Rao inequality \cite{Braunstein:1994, Paris:2009,Jacobs:2014} as
 \begin{equation}\label{CR}
 \left\langle\delta {\gamma ^2}\right\rangle \ge \frac{1}{{N\mathcal{I}\left( \gamma  \right)}},
\end{equation}
where $\mathcal{I}(\gamma)$ is the Fisher information and $N$ is the number of measurements. To improve the precision of the parameter estimation and to eliminate the decoherence effect as well, the Hamiltonian control method \cite{Hou:2017,Jacobs:2014,Yuan:2015,Egger:2014,Chen:2016,Zhang2012,Wu2015,Wu2013,Zhang2017,Qi2017,Zhang2010} can be used.
Here, we apply the linear control ${H_1} = {k_1}{a^\dag }a$ and Kerr-type nonlinear control \cite{Rossi:2016,Boyd:2008,Stobinska:2008,Milburn:1986} ${H_2} = {k_2}{\left( {{a^\dag }a} \right)^2}$ to  improve the precision of the estimation.
The Lindblad master equation~\eqref{Equ:Lindblad_Master_Equ} can be written as
\begin{equation}\label{Equ:Lindblad_ME_Linear_Nonlinear_Backaction}
%\frac{{d\rho }}{{d\tau }}=
\dot{\rho}=- i[u_1{a^\dag }a+u_2(a^\dag a)^2,\rho]+ ( {a\rho {a^\dag } - \frac{1}{2}{a^\dag }a\rho  - \frac{1}{2}\rho {a^\dag }a} ),
\end{equation}
where $\tau  = \gamma t,~{u_1} = {{{k_1}}}/{\gamma }$, and ${u_2} = {{{k_2}}}/{\gamma }$.
We assume that the initial state of the quantum system is a coherent state, ${\rho _0} = \left| \alpha  \right\rangle \left\langle \alpha  \right|$. The analytic solution of Eq.~\eqref{Equ:Lindblad_ME_Linear_Nonlinear_Backaction} in the interaction picture is
\begin{equation*}
\rho_I(\tau)=\sum_{l=0}^{\infty}\frac{\left(\frac{1-e^{-\Delta\tau}}{\Delta}\right)^n}{l!}\exp\Big\{-\frac{1}{2}\Delta\tau(p+q)\Big\}{a_I}^l\rho_{I_0}\left(a_I^{\dag}\right)^l
\end{equation*}
where, $a(\tau)=e^{-iu_1\tau a^{\dag}a}a_I(\tau)e^{iu_1\tau a^{\dag}a}$ and $\Delta  = 1 + 2i{u_2}\left( {p - q} \right)$.
%{\color{blue}In addition, ${\rho _{{I_0}}}$ can be approximately equal to ${\rho _0}$.}
Thus, the matrix elements $\rho_{p,q}$ of  Eq.~\eqref{Equ:Lindblad_ME_Linear_Nonlinear_Backaction}
 can be written as,
\begin{equation}\label{Equ:rho_Matrix_Element}
\begin{aligned}
{\rho _{p,q}}& = \lambda\exp \left\{ { - \frac{1}{2}\Delta \tau \left( {p + q} \right)} \right\}\cdot  \\
&\exp \bigg\{- i{u_1}\tau\left(p -q \right) - {{\left| \alpha  \right|}^2}\left( {1 - \frac{{1 - {e^{ - \Delta \tau }}}}{\Delta }} \right)\bigg\},\\
&~~~~~~~~~~~~~~~~~~~~~~~~~\text{for}~~~ p,~q = 0,~1,~2,~3,~\ldots
\end{aligned}
\end{equation}
where $\lambda=\alpha ^p {\bar \alpha }^q/ \sqrt {p!q!}$ and $\bar n = \left| \alpha  \right|^2$.

If $u_1 \ll 1$ and $u_2 \ll 1$, by using series expansions of the exponential in Eq.~\eqref{Equ:rho_Matrix_Element} for small $\tau$, the quantum state can be approximated as
\begin{equation}\label{Equ:rho_Pure_State_approx}
\begin{aligned}
 {\rho _{p,q}}&=\lambda\exp \left\{ \! { - \frac{1}{2}\tau \left( {p + q} \right) - {{\left| \alpha  \right|}^2}{e^{ - \tau }}} \right\}\exp \big\{\! -i u_1\tau(p\!-\!q)\\
 &~ { - i{u_2}\tau \left( {{p^2} - {q^2}} \right) - i{u_2}{{\left| \alpha  \right|}^2}{\tau ^2}\left( {p - q} \right)} \big\}.
 \end{aligned}
\end{equation}
The lowest order of the expansion can be rewritten as a pure state, {$\tilde{\rho} = \left| {{\psi _\gamma }} \right\rangle_{\text{app~app}}\left\langle {{\psi _\gamma }} \right|$},  where
\begin{equation}\label{pure state}
\begin{aligned}
\left| {{\psi _\gamma }} \right\rangle_{\text{app}} &= \exp \left\{ { - \frac{1}{2}{{\left| \alpha  \right|}^2}{e^{ - \tau }}} \right\}\cdot\\
&~~~~\left[ {\begin{matrix}
1\\
{\alpha \exp \left\{ { - \frac{1}{2}\tau  - i\left( {{u_1} + {u_2}} \right)\tau  - i{u_2}{\tau ^2}{{\left| \alpha  \right|}^2}} \right\}}
\end{matrix}} \right].
%,
%\\
%\left\langle {{\psi _\gamma }} \right| &= \exp \left\{ { - \frac{1}{2}{{\left| \alpha  \right|}^2}{e^{ - \tau }}} \right\}\times\\
%&~~~~~~~~\left[ {\begin{matrix}
%1&{\bar \alpha \exp \left\{ { - \frac{1}{2}\tau  + i\left( {{u_1} + {u_2}} \right)\tau  + i{u_2}{\tau ^2}{{\left| \alpha  \right|}^2}} \right\}}
%\end{matrix}} \right]
\end{aligned}
\end{equation}

\begin{figure}[H]
\begin{minipage}[T]{1\linewidth}
\centering
\includegraphics[width=3in]{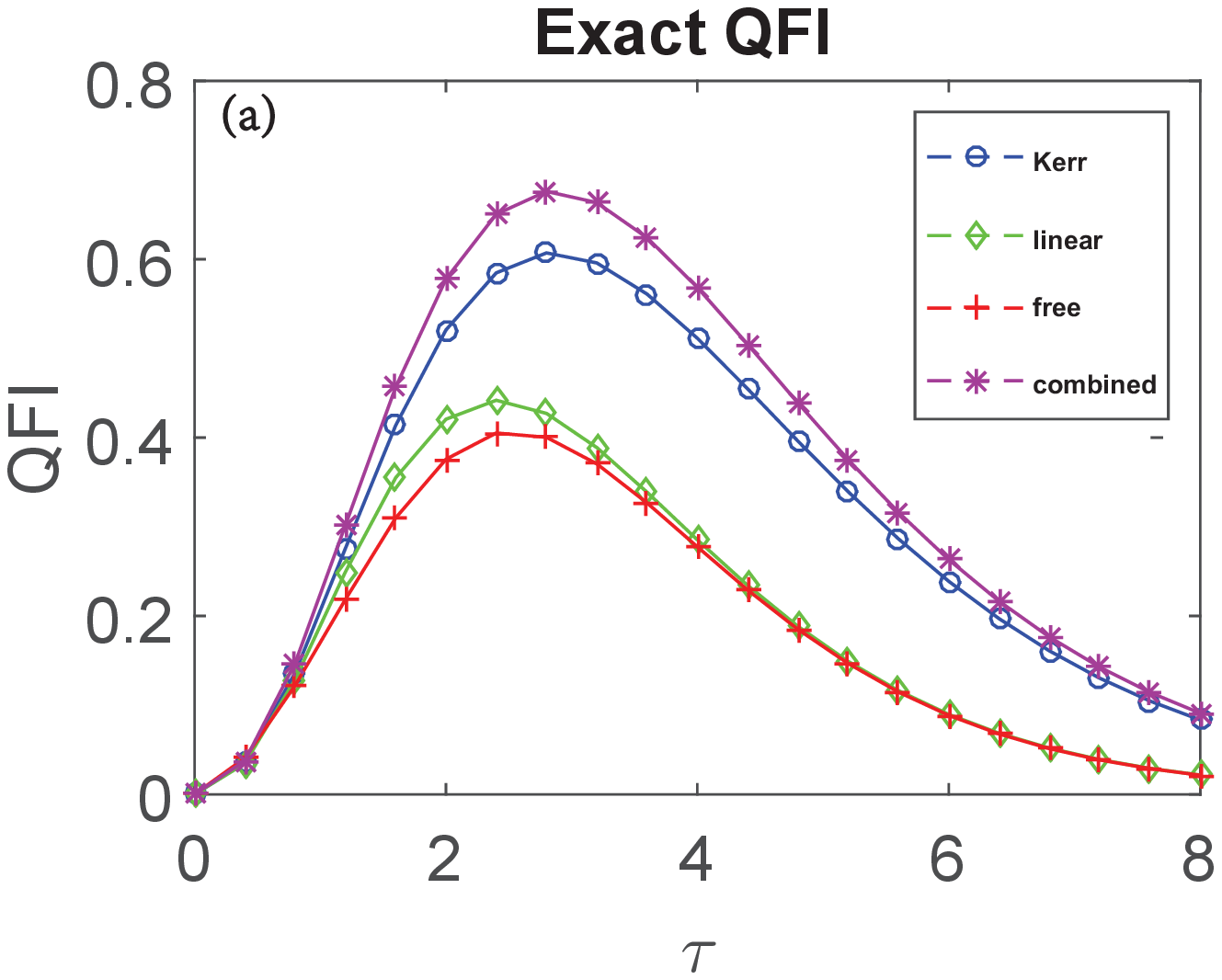}
\end{minipage}
\begin{minipage}[T]{1\linewidth}
\centering
\includegraphics[width=3in]{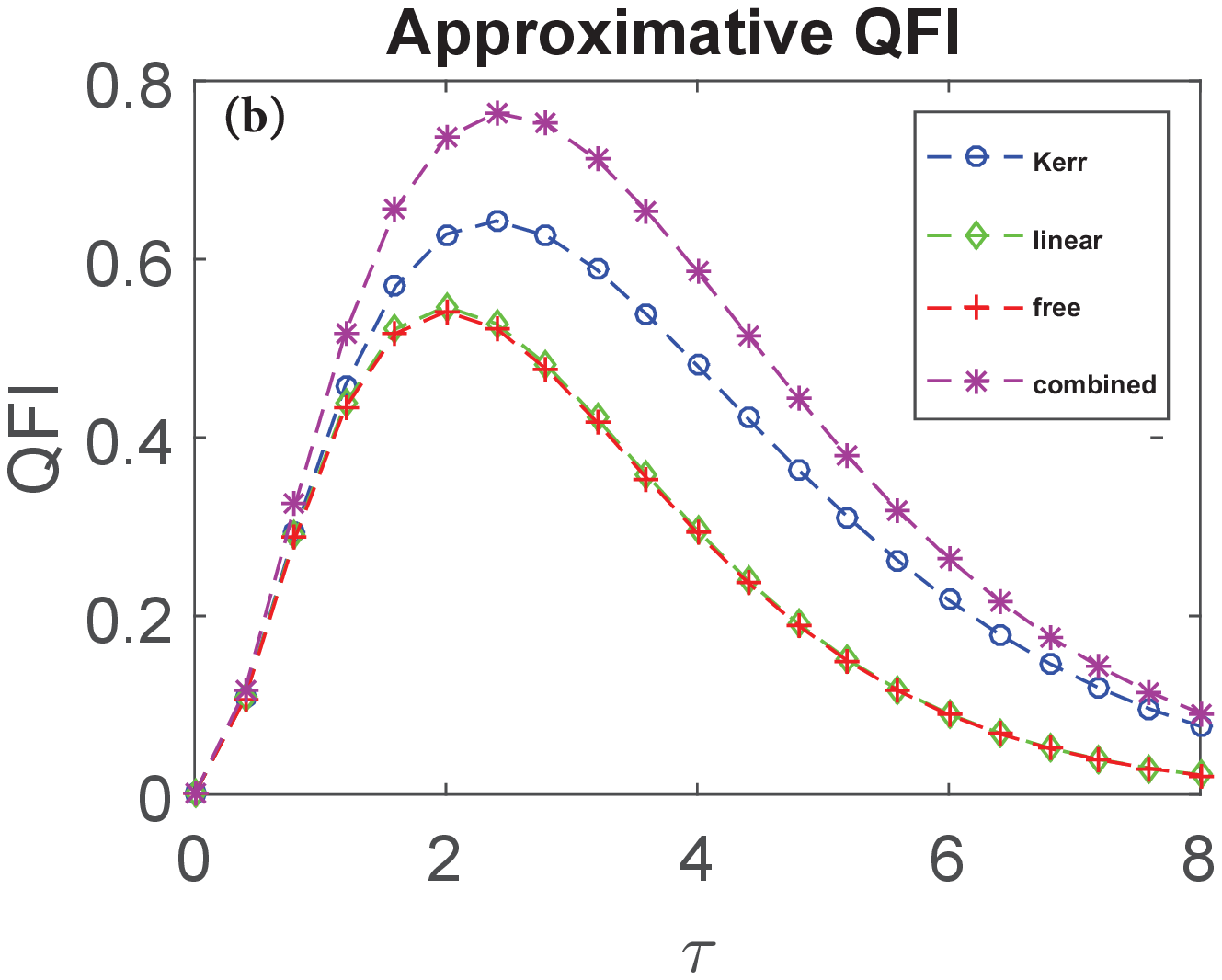}
\end{minipage}
\caption{(Color online) Evolution of the quantum Fisher information with various control Hamiltonians. The curves with red crosses represent the free evolution without control, the curves with green diamonds show the evolution with linear control $H_1$, the curves with blue circles show the evolution with Kerr-type nonlinear control $H_2$, and the curves with purple stars represent the evolution with both linear and Kerr-type nonlinear controls. {\color{red}Fig.~\ref{Fig:QFI}(a)} shows the evolution of the exact QFIs as functions of the rescaled time $\tau$, based on the first ten orders of the elements of the density matrix Eq.~\eqref{Equ:rho_Matrix_Element}. Based on the pure state approximation \eqref{pure state}, {\color{red}Fig.~\ref{Fig:QFI}(b)} shows the evolution of the approximative QFIs.
}\label{Fig:QFI}
\end{figure}

\section{Quantum parameter estimation}\label{sec:Quantum_parameter_estimation}
According to the quantum Cram\'{e}r$--$Rao inequality in Eq.~\eqref{CR}, the quantum Fisher information
 \begin{eqnarray}
 \mathcal{I}(\gamma) = \text{Tr}\left[ {{\rho }L_\gamma^2} \right]
 \end{eqnarray}
 provides an appropriate method to calculate the estimation precision. Here, ${L_\gamma }$, known as the
   \emph{system logarithmic derivative} \cite{Ciampini:2016} satisfies
 $${\partial _\gamma }{\rho } = \frac{{{L_\gamma }{\rho _\gamma } + {\rho _\gamma }{L_\gamma }}}{2}.$$
The diagonalization of density operator $\rho$ can simplify the quantum Fisher information (QFI), i.e,
if ${\rho  } = \sum\nolimits_n {{p_n}\left| {{\psi _n}} \right\rangle } \left\langle {{\psi _n}} \right|$ for some ${p_n},\left| {{\psi _n}} \right\rangle $, where $\sum\nolimits_n {{p_n}}=1$, the QFI can be written as
\begin{equation}\label{QFI2}
\mathcal{I}\left( \gamma  \right) = 2\sum\limits_{n,m} {\frac{{{{\left| {\left\langle {{\psi _m}} \right|{\partial _\gamma }{\rho _\gamma }\left| {{\psi _n}} \right\rangle } \right|}^2}}}{{{p_m} + {p_n}}}},
\end{equation}
where ${{p_m} + {p_n} \ne 0}$ for all $n$, $m$. Moreover, for a pure state ${\tilde{\rho} _\gamma}^2  = {\tilde{\rho} _\gamma }$, Eq.~\eqref{QFI2} can be rewritten as
\begin{equation}\label{QFI3}
\mathcal{I}\left( \gamma  \right) = 4\left[ {\left\langle {{\partial _\gamma }{\psi _\gamma }} \right|\left. {{\partial _\gamma }{\psi _\gamma }} \right\rangle  - {{\left( {\left\langle {{\psi _\gamma }} \right|\left. {{\partial _\gamma }{\psi _\gamma }} \right\rangle } \right)}^2}} \right].
\end{equation}
The partial differentiation over the pure state in Eq. \eqref{pure state}  with respect to $\gamma$ gives
%\begin{equation}
%\begin{aligned}
%\left| \partial_{\gamma}\psi _\gamma  \right\rangle_{\text{free}}  &= t\exp \left\{ { - \frac{1}{2}{{\left| \alpha  \right|}^2}{e^{ - \tau }}} \right\}\times\\
%& \left[ {\begin{matrix}
%\frac{1}{2}\big|\alpha\big|^2e^{-\tau}\\
%{\alpha \bigg(\frac{1}{2}|\alpha|^2e^{-\tau}-\frac{1}{2}\bigg) \exp \big\{  - \frac{1}{2}\tau \big\}}
%\end{matrix}} \right];
%\end{aligned}
%\end{equation}
%\begin{equation}
%\begin{aligned}
%&\left| \partial_{\gamma}\psi _\gamma  \right\rangle_{\text{linear}}  = t\exp \left\{ { - \frac{1}{2}{{\left| \alpha  \right|}^2}{e^{ - \tau }}} \right\}\times\\
%& \left[ {\begin{matrix}
%\frac{1}{2}\big|\alpha\big|^2e^{-\tau}\\
%{\alpha \bigg(\frac{1}{2}|\alpha|^2e^{-\tau}-\frac{1}{2}-iu_1\bigg) \exp \big\{  - \frac{1}{2}\tau -iu_1\tau\big\}}
%\end{matrix}} \right];
%\end{aligned}
%\end{equation}
%\begin{equation}
%\begin{aligned}
%&\left| \partial_{\gamma}\psi _\gamma  \right\rangle_{\text{Kerr}}  = t\exp \left\{ { - \frac{1}{2}{{\left| \alpha  \right|}^2}{e^{ - \tau }}} \right\}\times\\
%& \left[ {\begin{matrix}
%\frac{1}{2}\big|\alpha\big|^2e^{-\tau}\\
%{\alpha \bigg(\frac{1}{2}|\alpha|^2e^{-\tau}-\frac{1}{2}-iu_2-2iu_2|\alpha|^2\tau\bigg) \exp \big(M \big)}
%\end{matrix}} \right],
%\end{aligned}
%\end{equation}
%here $M=- \frac{1}{2}\tau -iu_2\tau-iu_2|\alpha|^2\tau^2 $; and
\begin{equation}
\begin{aligned}
&\left| \partial_{\gamma}\psi _\gamma  \right\rangle_{\text{app}}  = t\exp \left\{ { - \frac{1}{2}{{\left| \alpha  \right|}^2}{e^{ - \tau }}} \right\}\cdot\\
& \left[ {\begin{matrix}
\frac{1}{2}\big|\alpha\big|^2e^{-\tau}\\
{\alpha \bigg(\frac{1}{2}|\alpha|^2e^{-\tau}-\frac{1}{2}-i(u_1+u_2)-2iu_2|\alpha|^2\tau\bigg) \exp \big(N \big)}
\end{matrix}} \right],
\end{aligned}
\end{equation}
with $N=- \frac{1}{2}\tau -i(u_1+u_2)\tau-iu_2|\alpha|^2\tau^2$.
Thus, with the pure state approximation, the approximative QFI is
\begin{eqnarray}
\mathcal{I}_{\text{app}}(\tau)
&\approx &\frac{{{\tau ^2}}}{{{\gamma ^2}}}{\left| \alpha  \right|^2}{e^{{\rm{ - }}\tau }}\left( {1 + 4{{\left( {{u_1} + {u_2} + 2\tau {{\left| \alpha  \right|}^2}{u_2}} \right)}^2}} \right).
\end{eqnarray}
To confirm the validity of the pure state approximation, we plot the evolution of the QFI with various control Hamiltonians in {\color{red}Fig.~\ref{Fig:QFI}}.
\begin{figure}[H]
  \centering
  \includegraphics[width=8cm]{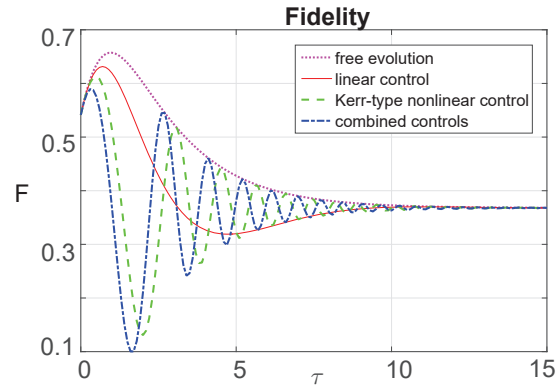}
  \caption{
Time evolution of fidelity with various control Hamiltonians. The purple dotted curve is the free evolution without control, the red solid curve represents the evolution with linear control, the green dashed curve indicates the evolution with Kerr-type nonlinear control, and the blue dash-dotted curve represents the evolution with both linear and Kerr-type nonlinear controls.
}\label{Fig:fidelity}
\end{figure}
Here, we choose the mean energy $\bar n=1$ and $u_1=u_2=0.05$.  The curves with red crosses represent the free evolution without control, the curves with green diamonds show the evolution with linear control $H_1$, the curves with blue circles show the evolution with Kerr-type nonlinear control $H_2$, and the curves with purple-star represent the evolution with both linear and Kerr-type nonlinear controls. {\color{red}Fig.~\ref{Fig:QFI}(a)} shows the evolution of the exact QFIs as functions of the rescaled time $\tau$, based on the first ten orders of the elements of the density matrix in Eq.~\eqref{Equ:rho_Matrix_Element}. Based on the pure state approximation in Eq. \eqref{pure state}, {\color{red}{Fig.~\ref{Fig:QFI}(b)}} presents the evolution of the approximative QFIs.
The comparison between the two figures shows the effectiveness of the pure state approximation. Thus, we optimize the control input based on the pure state approximated QFI.

\begin{figure}[H]
\begin{minipage}[H]{1.0\linewidth}
\centering
\includegraphics[width=3in]{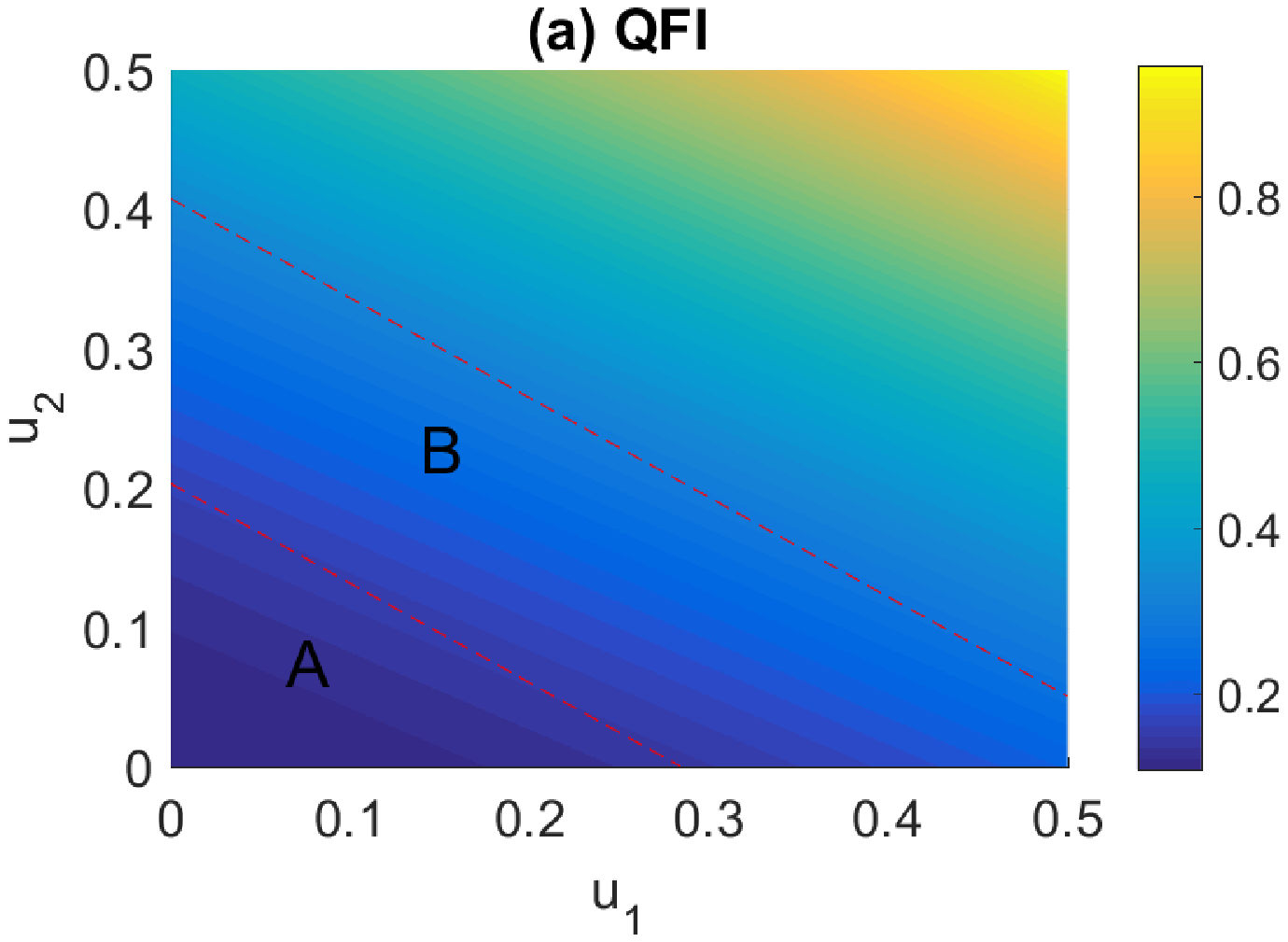}
\end{minipage}
\begin{minipage}[H]{1.0\linewidth}
\centering
\includegraphics[width=3in]{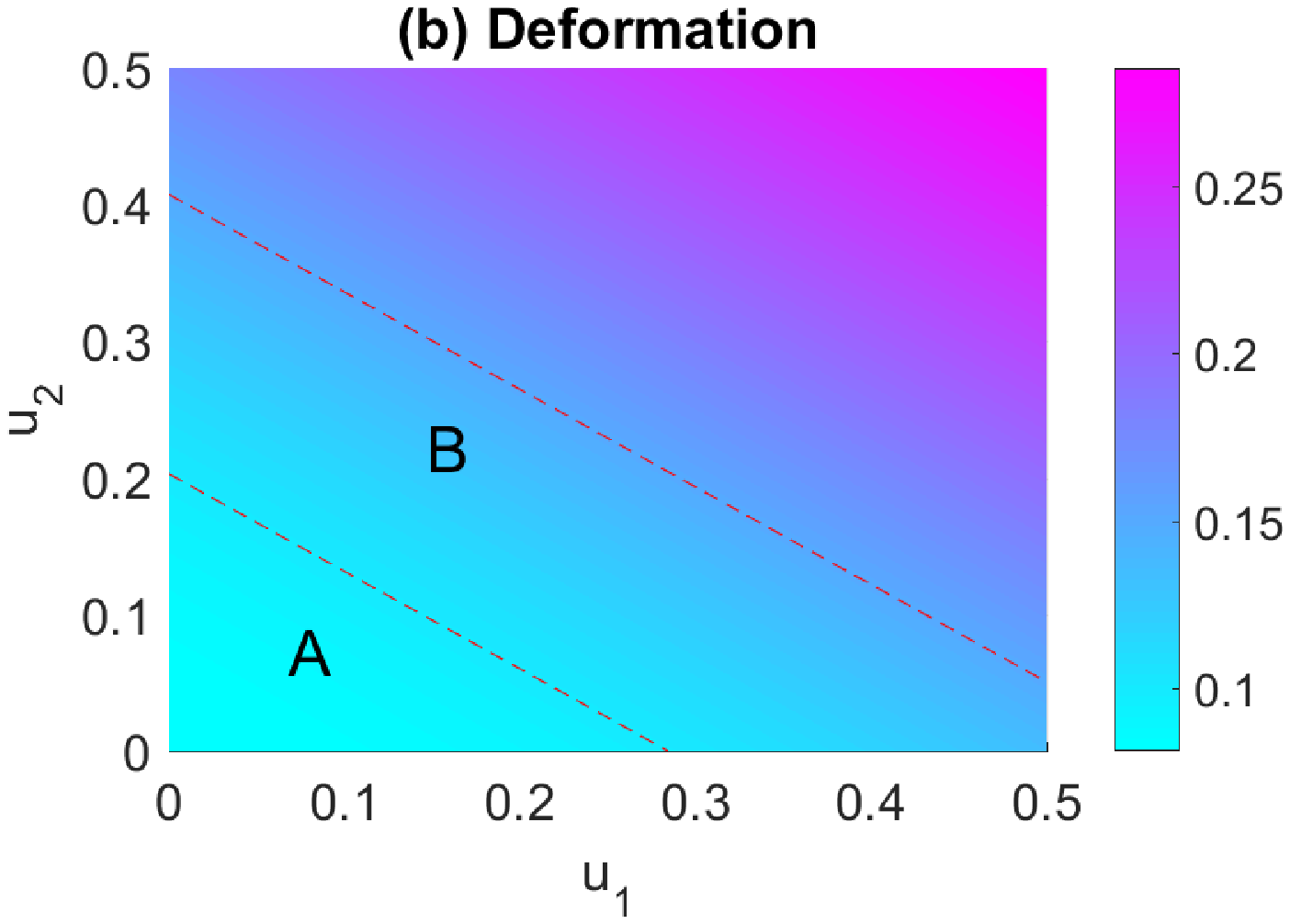}
\end{minipage}
\caption{(Color online)\label{Fig:optimization}
Illustration of the $\varepsilon$-constrained multi-objective optimization for quantum parameter estimation. In {\color{red}Fig.~\ref{Fig:optimization}(a)} and {\color{red}Fig.~\ref{Fig:optimization}(b)}, the Fisher information $\mathcal{I}_\text{app}^{\ast}$ and the deformation $\mathcal{D}$ as functions of linear and Kerr-type nonlinear controls are plotted with $|\alpha|^2=0.2$. In {\color{red}Fig.~\ref{Fig:optimization}(b)}, region A and region B indicate the constraints $\varepsilon=0.10$ and $\varepsilon=0.15$, respectively. Thus, the $\varepsilon$-constrained optimum value of the quantum Fisher information is restricted in A or B in {\color{red}Fig.~\ref{Fig:optimization}(a)}.
}
\end{figure}
In the following, we calculate the deformation of the system state
 during the parameter estimation process by the quantum fidelity,  $F\left( {{\rho _0},\rho } \right) = {\left( {{\rm Tr}\left( {\sqrt {\sqrt {{\rho }} \rho_0 \sqrt {{\rho}} } } \right)} \right)^2}$. It is obvious that the fidelity is symmetric with respect to $\rho_0$ and $\rho$, and ranges from 0 to 1. If {${\rho _{\rm{0}}}{\rm{ = }}\left| \varphi  \right\rangle \left\langle \varphi  \right|,\rho {\rm{ = }}\left| \psi  \right\rangle \left\langle \psi  \right|$} are pure states, the fidelity is
reduced to a relatively simple form of $F=\left| {\left\langle \psi  \right|\left. \varphi  \right\rangle } \right|^2$. Thus, based on Eq.~\eqref{pure state} the quantum state fidelity can be written as
\begin{eqnarray}\label{Equ:Fidelity_rho_rho_0}
F(\tau)&=& \bigg| \exp \left\{ { - \frac{1}{2}{{\left| \alpha  \right|}^2} - \frac{1}{2}{{\left| \alpha  \right|}^2}{e^{ - \tau }}} \right\} \bigg[ 1 +\nonumber\\
 &&{{\left| \alpha  \right|}^2}\exp \left\{ { - \frac{1}{2}\tau  + i\left( {{u_1} + {u_2}} \right)\tau  + i{u_2}{\tau ^2}{{\left| \alpha  \right|}^2}} \right\} \bigg] \bigg|^2.
\end{eqnarray}
In {\color{red}Fig.~\ref{Fig:fidelity}}, the evolution of the fidelity is plotted as a function of the rescaled time $\tau$ with various control Hamiltonians. The purple dotted curve is the free fidelity evolution without control, the red solid curve represents the fidelity evolution with linear control, the green dashed curve indicates the fidelity evolution with Kerr-type nonlinear control, and the blue dash-dotted curve represents the fidelity evolution with both linear and Kerr-type nonlinear controls. The figure shows that while the parameter estimation precision is improved, it introduces a significant deformation to the system state.

\section{Multi-objective optimization}\label{sec:Multi-objective_optimization}
The results in the above section show a clear trade-off relation between the parameter estimation precision and the fidelity. Here, we have two conflicting objectives that require optimization. In Ref. \cite{Cui:2012}, the second author and co-authors of the article used goal programming to deal with a similar problem in a quantum state reconstruction. Instead of finding solutions, which can absolutely
minimize or maximize objective functions, the function of
goal
programming is to find solutions that, if possible, satisfy a set of goals, or otherwise violate the goals minimally. In this study, we use the multi-objective optimization method \cite{Ehrgott:2006,Miettinen:1999} to deal with this problem.
The multi-objective optimization problem can be formulated as
\begin{eqnarray}
&&
{\rm Minimize}\quad
\left( {{f_1}\left( x \right),{f_2}\left( x \right), \ldots ,{f_k}\left( x \right)} \right)
 \nonumber \\
&&
{\rm subject\,\,to}\,~~~x \in X,
\end{eqnarray}
where there are $k~(\geq2)$ objective functions $\left\{ {{f_1},{f_2},...{f_k}} \right\}$. The decision vector $x = {\left[ {{x_1},{x_2}, \cdots ,{x_n}} \right]^T}$ belongs to a feasible set of $X$.
Because of the contradiction of the objectives ${{f_1}\left( x \right),{f_2}\left( x \right), \ldots ,{f_k}\left( x \right)} $, it is not possible to find a single solution, which is optimal for all the objectives simultaneously.
The common solution is the so-called Pareto front, where there is no other solution that dominates it \cite{Ehrgott:2006,Miettinen:1999}. The solution set can be nonconvex and nonconnected. A feasibly efficient way is to formulate a single-objective optimization problem, and the optimal solutions to the single-objective optimization problem are Pareto optimal solutions to the original multi-objective optimization problem. The $\varepsilon$-constraint is a widely used method
\begin{eqnarray}\label{optimization2}
&&
{\rm Minimize}\quad
f_j(x)
 \nonumber \\
&&
{\rm subject\,\,to}\,
\left\{
\begin{array}{l}
x \in X \\
f_i(x)\leq\varepsilon_j~~\text{for}~~i\in\{1,\cdots,k\}\setminus\{j\}.
\end{array}
\right.
\end{eqnarray}
In our optimization problem, the decision vector is defined as $x = {[ {\tau ,{u_1},{u_2},|\alpha {|^2}} ]^T}$. In order to simplify the discussion, let us take the partial derivative of the quantum  Fisher information with respect to $\tau$ and let the results be $0$. This is a reasonable assumption because in the quantum parameter estimation precision the maximum value of the Fisher information is of particular interest. Thus, we obtain the maximum value $\mathcal{I}_\text{app}^{\ast}$ as:
\begin{eqnarray}
%\mathcal{I}_\text{free}^{\ast}&\approx&\frac{4\left|\alpha\right|^2}{\gamma^2}e^{-2},\\
%\mathcal{I}_\text{linear}^{\ast}&\approx&\frac{4\left|\alpha\right|^2}{\gamma^2}e^{-2}\bigg(1+4u_1^2\bigg),\\
%\mathcal{I}_\text{Kerr}^{\ast}&\approx &\frac{4\left|\alpha\right|^2}{\gamma^2}e^{-2}\bigg(1+64u_2^2\left|\alpha\right|^2\bigg),\\
\mathcal{I}_\text{app}^{\ast}&\approx & \frac{{4{{\left| \alpha  \right|}^2}}}{{{\gamma ^2}}}e^{-2}\left( {1 + 4{{\left( {{u_1} + {u_2} + 4{{\left| \alpha  \right|}^2}{u_2}} \right)}^2}} \right).
\end{eqnarray}
Obviously, the extremum point  $\tau^{\ast}$ is the solution of the following equations,
\begin{eqnarray}
-16u_2^2\big|\alpha\big|^4{\tau^{\ast}}^3+64u_2^2\big|\alpha\big|^4{\tau^{\ast}}^2-\tau^{\ast}+2=0,\\
-4T^2\tau^{\ast}+8T^2+16\big|\alpha\big|^2u_2T\tau^{\ast}-\tau^{\ast}+2=0, \\
\text{with}~~T=u_1+u_2+2\big|\alpha\big|^2u_2\tau^{\ast}.\nonumber
\end{eqnarray}
 Because $u_1 \ll 1$, $u_2 \ll 1$, the approximate solution of the above equations is $\tau^{\ast}=2$.  It is clear that the quantum Fisher information is a monotonically increasing function of the linear and Kerr-type nonlinear controls, $u_1$, $u_2$, and $|\alpha|^2$.
 Nevertheless, at the same time the deformation to the quantum state (for a given Gaussian state as an input, over time the state becomes non-Gaussian) is
 \begin{eqnarray}\label{B}
 \mathcal{D} &=&1-F\big|_{\tau=\tau^{\ast}} \nonumber\\
 &\approx& 1 - \bigg| \exp \left\{ { - \frac{1}{2}{{\left| \alpha  \right|}^2} - \frac{1}{2}{{\left| \alpha  \right|}^2}{e^{ - 2}}} \right\}\cdot \\
 &&\left[ { 1 + {{\left| \alpha  \right|}^2}\exp \left\{ {-1+2i\left( {{u_1} + {u_2}} \right) + 4i{u_2}{{\left| \alpha  \right|}^2}} \right\}} \right] \bigg|^2. \nonumber
 \end{eqnarray}
Finally, the $\varepsilon$-constrained multi-objective optimization problem can be written as,
\begin{eqnarray}\label{optimization3}
&&
{\rm Maximum}\quad
\mathcal{I}_{\text{app}}^{\ast}
 \nonumber \\
&&
{\rm subject\,\,to}\,~~
\left\{
\begin{array}{l}
\mathcal{D}\leq\varepsilon\\
u_1,~u_2,~|\alpha|^2 \in [0,1).
\end{array}
\right.
\end{eqnarray}
The parameter $\varepsilon$ is regarded as the permissible damage of the initial state.

\begin{figure}[H]
\begin{minipage}[H]{1.0\linewidth}
\centering
\includegraphics[width=3.0in]{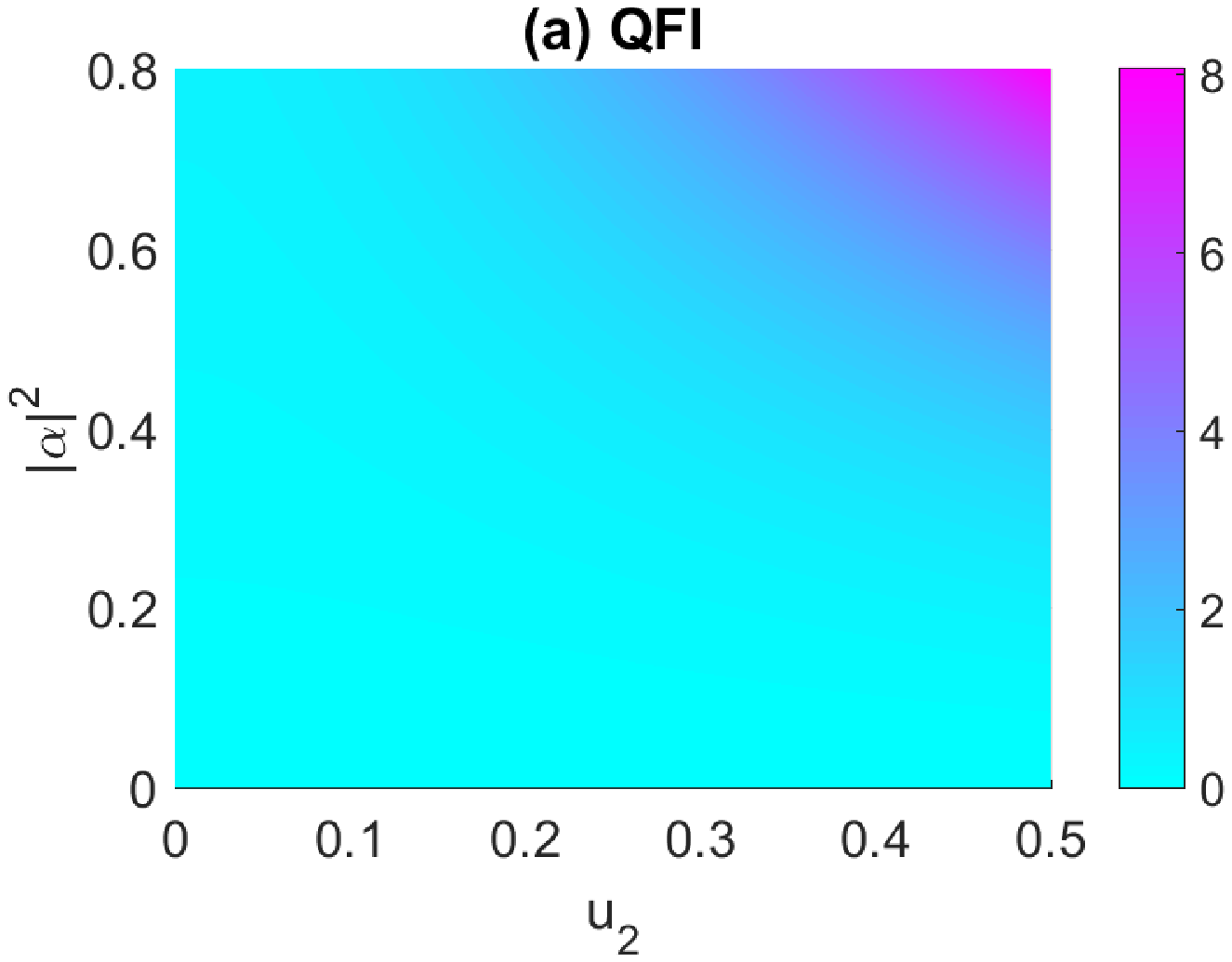}
\end{minipage}
\begin{minipage}[H]{1.0\linewidth}
\centering
\includegraphics[width=3.0in]{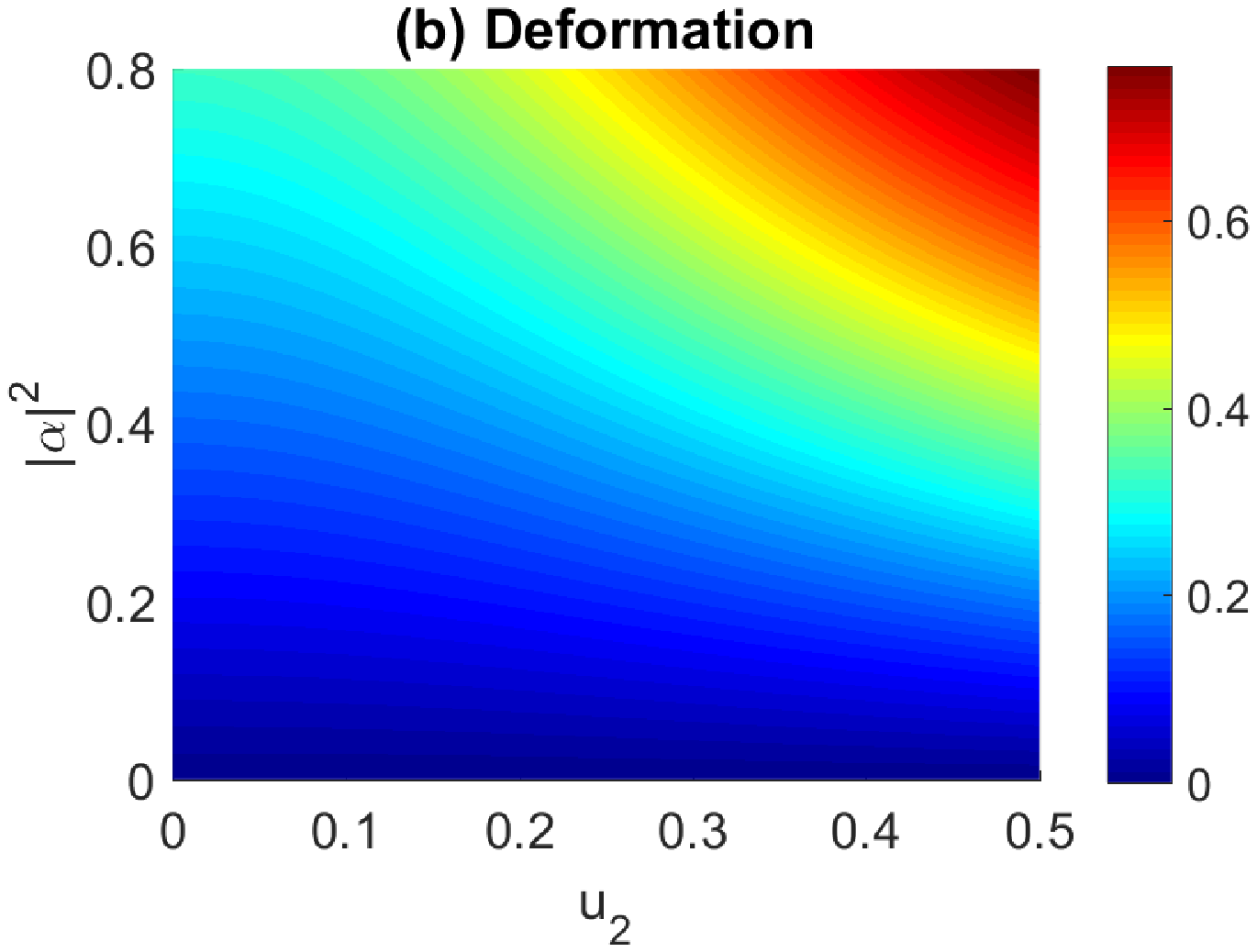}
\end{minipage}
\caption{(Color online)
Evolution of the Fisher information $\mathcal{I}_\text{app}^{\ast}$ and the deformation $\mathcal{D}$ as functions of the Kerr-type nonlinear control and the parameter $|\alpha|^2$.
}\label{Fig:optimization1}
\end{figure}

In {\color{red}Fig.~\ref{Fig:optimization}}, the quantum Fisher information $\mathcal{I}_\text{app}^{\ast}$ and the deformation $\mathcal{D}=1-F$ as functions of the linear control $u_1$ and Kerr-type nonlinear control $u_2$ with $|\alpha|^2=0.2$ can be seen.  By choosing the permissible damage $\varepsilon$, a control region $u_1\times u_2$ is generated, as shown in {\color{red}Fig.~\ref{Fig:optimization}(b)}. Subsequently, in this region, the quantum Fisher information can be optimized, as shown in {\color{red}Fig.~\ref{Fig:optimization}(a)}.
An example is shown in {\color{red}Fig.~\ref{Fig:optimization}(b)}, where region A and region B indicate the constraints $\varepsilon=0.10$ and $\varepsilon=0.15$, respectively. Thus, the $\varepsilon$-constrained optimum value of the quantum Fisher information is restricted in A or B in {\color{red}Fig.~\ref{Fig:optimization}(a)}.  When information from the measurement is acquired and the QFI is improved, it obviously introduces a significant back action to the system itself and destroy the quantum state.
As it is shown above, the Fisher information is a monotonically increasing function of the linear and Kerr-type nonlinear controls. Thus, the $\varepsilon$-constrained optimum values of the quantum Fisher information are located in the front dashed lines.  To gain more insight into the actual actions during the estimation process, we consider the trade-off between QFI and the deformation as functions of other parameters.
The case where only Kerr-type nonlinear control is applied is shown in {\color{red}Fig.~\ref{Fig:optimization1}}.
It is clear that the parameter $|\alpha|^2$ also plays an important role in the optimization of the Fisher information in the quantum parameter estimation.  As a preliminary work, this paper only considers the condition under the evolution of a single parameter, while the studies of the collaborative optimization is the subject of further research.

 In this paper, we studied a multi-objective optimization problem to improve the precision of parameter estimation under the constraint of state fidelity. However, the method of estimating the parameter by practical measurement records is of particular interest. In the following, we summarize the process of estimating the dissipation rate $\gamma$ in a stochastic master equation (SME) with a quantum weak measurement:
i) Generate a series of choices $\gamma_1$, $\gamma_2$, $\cdots$, with certain probability;
ii) Solve the SME and obtain the trajectory of the state $\rho_{\gamma_i}$ for each $\gamma_i,~~(i=1,2,\cdots n)$. The ensemble state $\rho$ is calculated by $\bar{\gamma}=\frac{1}{n}\sum_{i=1}^n\gamma_i$;
iii) Obtain $P_i(t)$ and calculate the estimator $\hat{\gamma}$ by the formula $\hat{gamma}(t)=\sum_{i=1}^n\gamma_iP_i(t)$.

\section{Conclusion}\label{sec:Conclusion}

In several realistic quantum systems, the reliability of quantum parameter estimation is an important issue. In this paper,
we used the linear and Kerr-type nonlinear controllers to improve the precision of the estimation of unknown parameters that govern the system evolution. In particular, the estimation of the dissipation rate of a quantum master equation has been considered. We show that while the precision of the parameter estimation is improved, it usually introduces significant deformation to the system state.
We propose a multi-objective model to maximize the Fisher information, as well as to minimize the deformation on the quantum system. Finally, simulations of a simplified $\varepsilon$-constrained model demonstrated the feasibility of the Hamiltonian control in the quantum parameter estimation.
The simulation results demonstrated the feasibility of
Hamiltonian control in improving the precision of the quantum parameter estimation.

%%%%%%%%%%%%%%%%%%%%%%%%%%%%%%%%%%%%%%%%%%%%%%%%%%%%%%%
%%% Acknowledgements.
%%%%%%%%%%%%%%%%%%%%%%%%%%%%%%%%%%%%%%%%%%%%%%%%%%%%%%%
\Acknowledgements{This work was supported by the National Natural Science Foundation of China, under Grant 11404113,
and the Guangzhou Key Laboratory of Brain Computer Interaction and Applications, under Grant 201509010006.}

%%%%%%%%%%%%%%%%%%%%%%%%%%%%%%%%%%%%%%%%%%%%%%%%%%%%%%%
%%% Conflict of interest. ????????????
%%%%%%%%%%%%%%%%%%%%%%%%%%%%%%%%%%%%%%%%%%%%%%%%%%%%%%%
\InterestConflict{The authors declare that they have no conflict of interest.}

%%%%%%%%%%%%%%%%%%%%%%%%%%%%%%%%%%%%%%%%%%%%%%%%%%%%%%%
%%% Supplements. ????????, ????
%%%%%%%%%%%%%%%%%%%%%%%%%%%%%%%%%%%%%%%%%%%%%%%%%%%%%%%
%\Supplements{}

%%%%%%%%%%%%%%%%%%%%%%%%%%%%%%%%%%%%%%%%%%%%%%%%%%%%%%%
%%% Reference section. ?????
%%% citation in the content using "some words~\cite{1,2}".
%%% ~ is needed to make the reference number is on the same line with the word before it.
%%%%%%%%%%%%%%%%%%%%%%%%%%%%%%%%%%%%%%%%%%%%%%%%%%%%%%%

%%%%%%%%%%%%%%%%%%%%%%%%%%%%%%%%%%%%%%%%%%%%%%%%%%%%%%%
%%% Appendix sections. ??????, ????
%%%%%%%%%%%%%%%%%%%%%%%%%%%%%%%%%%%%%%%%%%%%%%%%%%%%%%%

\end{multicols}
\end{document}